# Orthogonal Persistence Revisited

Alan Dearle, Graham N.C. Kirby and Ron Morrison

School of Computer Science, University of St Andrews,
North Haugh, St Andrews, Fife KY16 9SX, Scotland
{al, graham, ron}@cs.st-andrews.ac.uk

**Abstract.** The social and economic importance of large bodies of programs and data that are potentially long-lived has attracted much attention in the commercial and research communities. Here we concentrate on a set of methodologies and technologies called persistent programming. In particular we review programming language support for the concept of orthogonal persistence, a technique for the *uniform* treatment of objects irrespective of their types or longevity. While research in persistent programming has become unfashionable, we show how the concept is beginning to appear as a major component of modern systems. We relate these attempts to the original principles of orthogonal persistence and give a few hints about how the concept may be utilised in the future.

## 1 Introduction

The aim of persistent programming is to support the design, construction, maintenance and operation of long-lived, concurrently accessed and potentially large bodies of data and programs. When research into persistent programming began, persistent application systems were supported by disparate mechanisms, each based upon different philosophical assumptions and implementation technologies [1]. The mix of technologies typically included naming, type and binding schemes combined with different database systems, storage architectures and query languages.

The incoherence in these technologies increased the cost both intellectually and mechanically of building persistent application systems. The complexity distracted the application builder from the task in hand to concentrate on mastering the multiplicity of programming systems, and the mappings amongst them, rather than the application being developed. The plethora of disparate mechanisms was also costly in machine terms, in that the code for interfacing them, their redundant duplication of facilities and their contention for resources caused execution overheads. Software architects and engineers observed that it was often much harder and more expensive to build and maintain persistent application systems than was expected, and their evolution was invariably problematic.

Atkinson [2] postulated that, in many cases, the inconsistency was not fundamental but accidental. The various subsystems were built at different times when the engineering trade-offs were different. In consequence, they provided virtually the same services, but inconsistently since they were designed and developed independently. By contrast, Orthogonal Persistence provided the total composition of services within one coherent design, thereby eliminating these accidental disharmonies.

While research in persistent programming has become unfashionable, it is hard to believe that the situation today has changed much. A recent (2007) quote from Microsoft illustrates this well:

> "*Most programs written today manipulate data in one way or another and often this data is stored in a relational database. Yet there is a huge divide between modern programming languages and databases in how they represent and manipulate information. This impedance mismatch is visible in multiple ways. Most notable is that programming languages access information in databases through APIs that require queries to be specified as text strings. These queries are significant portions of the program logic. Yet they are opaque to the language, unable to benefit from compile-time verification and design-time features like IntelliSense.*" [3]

Orthogonally persistent object systems support a *uniform* treatment of objects irrespective of their types by allowing values of *all* types to have whatever longevity is required. The benefits of orthogonal persistence have been described extensively in the literature [2,4-18]. They can be summarised as:

- improving programming productivity from simpler semantics;
- avoiding ad hoc arrangements for data translation and long-term data storage;
- providing protection mechanisms over the whole environment;
- supporting incremental evolution; and
- automatically preserving referential integrity over the entire computational environment for the whole life-time of an application.

In this paper we review a selection of the many historical approaches to programming with long-lived data[1] and comment on attempts in the programming language and ODBMS communities to provide various flavours of persistence. We conclude by hinting at how the concept may be utilised in the future.

## 2 Orthogonal Persistence

In most current application systems there are two domains: the programming language domain and the database domain. The programming language domain presents a Turing-complete programming environment that permits computation over data defined using the programming language type system. In the last twenty years the predominant programming model has become the object-oriented model, usually providing typed objects containing state, methods and (usually typed) references to other objects. This model, and the tools which have evolved to support it, has proven to be highly productive in terms of creating and maintaining software.

[1] Space limitations preclude a full survey of the area; notable omissions include Smalltalk, O2, Galileo, Trellis/Owl, Fibonacci, DBPL and Tycoon.

By contrast, the conceptual database domain is largely unchanged: tables of tuples containing foreign keys identifying tuples in other tables. This remains the pre-eminent long-term storage architecture.

The cost of the conceptual and technological differences between these two models became known as the *impedance mismatch* [19], and was one of the primary motivations for the work on orthogonal persistence, which aimed to remove the conceptually unnecessary distinction between short-term and long-term data [1].

There is a spectrum of possible degrees of integration, as perceived by programmers, between these formats. At one end of this spectrum data formats are completely disparate, and there is no automated support for transformation between them. A programmer has to understand the semantics of multiple representations and the mappings between them, and to write code for data transformations that implement these mappings. The impedance mismatch is strongest at this end of the spectrum. On the other hand, the low degree of integration yields loose coupling between the language and storage domains, which in turn facilitates openness in terms of the persistent data being accessible by routes other than the language.

At the full integration end of the spectrum lies orthogonal persistence, where no distinction between data formats is visible to the programmer. At intermediate points in the spectrum, the mapping between the object and storage domains is partially automated. Typically, the programmer still has to specify the mappings and understand the relationships between the multiple representations, but is relieved of the task of writing explicit translation code.

These differences are crystallised by Fowler, who describes two different architectural patterns that may be applied to persistent systems [20]. These are the *Active Record* and *Data Mapper* patterns. In the first, an object in a programming system represents a row in a database relation. In this pattern the database is wrapped in an object that provides methods to save, update, delete and find objects. Here there is a one-to-one mapping between classes or types in the programming language and relations in the database.

The *Data Mapper* pattern is more general. It comprises (potentially multiple) mappers that move data between the storage layers and maintains the relationships between entities. For example, in an object-relational system there is one mapper and two layers—the language system and the relational database. In a distributed system with caching there might be two mappers maintaining relationships between three layers—the language, the cache and the database.

The degree of integration dictates the extent to which the application programmer must be conscious of these patterns. With orthogonal persistence they are handled entirely by the system. Atkinson and Morrison identified three Principles of Orthogonal Persistence [21]:

- **The Principle of Persistence Independence**
  The persistence of data is independent of how the program manipulates the data. That is, the programmer does not have to, indeed cannot, program to control the movement of data between long term and short term store. This is performed automatically by the system.

- **The Principle of Data Type Orthogonality**
  All data objects should be allowed the full range of persistence irrespective of their type. That is, there are no special cases where objects of a specific type are not allowed to be persistent.

- **The Principle of Persistence Identification**
  The choice of how to identify and provide persistent objects is orthogonal to the universe of discourse of the system[2].

The application of the three principles yields orthogonal persistence. Violation of any of these principles increases the complexity that persistent systems seek to avoid. In the next section we examine these principles in the context of past and current persistent systems.

# 3 Languages and Persistence

## 3.1 First Generation Persistence Mechanisms

In the last twenty to thirty years the mechanisms for mapping between the two programming language and database data models have improved considerably. Ironically, this is in part due to technologies that were developed in the typed persistent world, for example strongly typed generative and reflective programming.

In the eighties it was common for programmers to explicitly save and restore programming language objects to the file system. Code was hand-written and tended to be error-prone and time consuming. Furthermore, when the data was changed, the code had to adapt, and more code written to evolve any saved data from previous program incarnations. The need to write such code explicitly was first eliminated by persistent systems such as PS-algol (discussed in the next section) and object-oriented databases.

Java serialization goes some way to reducing the programming effort required to implement object persistence using files, since it allows an entire closure to be written or read in a single operation. Only instances of classes that implement the interface *java.io.serializable* may be serialized. For example:

[2] Experience with persistent programming showed that in systems with references, the only mechanism for implementation was persistence by reachability, also known as transitive persistence.

```
FileInputStream f =
    new FileInputStream("myobject.data");
ObjectInputStream obj_in = new ObjectInputStream(f);
Object obj = obj_in.readObject ();
if (obj instanceof Person) {
    Person p = (Person) obj;
   // Do something with p ...
}
```

The above program reads an object from the file *myobject.data* and casts it to the type *Person*. One problem with this style of programming is that the entire closure of an object must be loaded or saved in a single operation. This can make the operations slow for large object closures, and limits the size of closure that can be stored to that of main memory—known as the *big inhale* in early Smalltalk-80 systems. However, more importantly, each time a closure is serialized a new copy of the data is made. This breaks referential integrity since there is no way of matching the identity of objects from different save/load operations. Another problem with the mechanism is that since not all Java classes are *serializable*, some object closures are not consistently saved and restored.

Serialization does not adhere to the first two principles of orthogonal persistence. Data is explicitly written to backing store, violating the principle of independence; only serializable objects may be made persistent, so the principle of data type orthogonality is also violated.

In contrast, to extract data from a database, the programs manipulating persistent data had to perform much string processing. Despite this approach manifesting a high impedance mismatch, it is still common in today's PHP programs. For example:

```
$result = mysql_query("SELECT * FROM Persons");
while($row = mysql_fetch_array($result)) {
    $firstname = $row['FirstName'];
    $secondname = $row['LastName'];
}
```

In this fragment [22], the database access is explicit—the SQL query is embedded as a string in the program, and data is extracted from the database in the form of strings.

The use of strings is also employed in JDBC [23], which provides database independent connectivity between Java programs and databases. The JDBC API permits SQL operations to be performed, by providing three broad classes of operations: establishing connections to a database, performing queries and processing the results of queries. An example is shown below:

```
Connection con = DriverManager.getConnection(
    "jdbc:myDriver:fish", "myLogin", "myPassword");
Statement stmt = con.createStatement();
ResultSet rs = stmt.executeQuery(
    "SELECT name,age FROM Persons");
while (rs.next()) {
    String name = rs.getString("name");
    int age = rs.getInt("age");
    ..
}
```

The similarities between the JDBC and PHP examples are striking. Both embed a query in the form of a string in the host program, and both use string matching to extract data from the result set that is delivered by the query. Both mechanisms are a long way from the principles of orthogonal persistence.

## 3.2 PS-algol

The first language to provide orthogonal persistence was PS-algol [1], which provided persistence by reachability for all data types supported by the language. PS-algol adds a small number of functions to S-algol [24], from which it was derived. These are *open.database*, *close.database*, *commit* and *abort*[3]. A number of functions are also provided to manage associative stores (hash maps), called *tables* in PS-algol. These functions are *s.lookup*, which retrieves a value associated with a key in a table, and *s.enter*, which creates an association between a key and a value in a table. By convention, a database always contains a pointer to a table at its root. Databases serve as roots of persistence and can be created dynamically.

Two slightly modified examples from [1] are shown below to give a flavour of the language. The first example opens a database called "addr.db" and places a person object into a table associated with the key "addr.table" found at its root. Note that the person (denoted by *p*) contains a reference to an address object. When *commit* is called, the updated table, the person and the address objects are written to persistent storage.

```
structure person(string name, phone; pntr addr)
structure address(int no ; string street, town)
let db = open.database("addr.db", "write")
if db is error.record
    do { write "Can't open database"; abort }
let table = s.lookup("addr.table", db)
let p = person("al", 3250,
               address(76, "North St", "St Andrews"))
s.enter("al", table, p)
commit
```

The second example opens the same database and retrieves the person object before writing out their phone number.

[3] Note: dots are legal within identifiers in PS-algol and do not denote dereferencing. Dereferencing is represented by round brackets enclosing a fieldname. There is no explicit *new* operator; the use of a structure name serves as a constructor.

```
structure person (string name, phone; pntr addr)
let db = open.database("addr.db", "read")
if db is error.record
   do { write "Can't open database"; abort }
let table = s.lookup("addr.table", db)
let p = s.lookup("al", table)
if p = nil then write "Person not known"
else write "phone number: ", p(phone)
```

As described in [25], “the programmer never explicitly organises data movement but it occurs automatically when data is used”, a feature shared with many of the object-relational systems. The paper also states “the language type rules are strictly enforced” but is not explicit about how this is achieved, which is a pity, since it is important. PS-algol uses structural type equivalence rather than the name equivalence so prevalent today. Using structural type equivalence, two objects or terms are considered to have compatible types if the types have identical structure. Thus, in the previous examples, the compatible declarations of *person* in the two examples serve to unify the two programs. If the object retrieved from the database is not of (structural) type *person*, the deference of the object will fail.

The type system of PS-algol is more subtle than might appear. Notice that the second program does not require a declaration of the type *address* since that type is never used in the program. It is not necessary since pointers in PS-algol are typed as *pntr*, which is an infinite union over all records. The infinite union facilitates partial and incremental specification of the structure of the data at the expense of a dynamic check. The persistent schema need only be specified within a program up to limit of the *pntr* objects. When one is encountered in a running program, by dereference, a dynamic check ensures the data is of the correct type. The specification within that check need only be to the limit of the subsequent *pntr* types.

A second version of PS-algol incorporated procedures as data objects thereby allowing code and data to be stored in the persistent store.

PS-algol does not support any form of concurrency other than at database level. This often caused problems since it was possible to continue to access objects after *commit*. The addition of explicit syntactic boundaries to control transactions would have addressed this deficiency.

## 3.3 Napier88

Napier88 attempted to explore the limits of orthogonal persistence by incorporating the entire language support environment within a strongly typed persistent store [12,21,26-30]. The research produced the first integrated, self-contained, type-safe persistent environment.

The Napier88 system provides orthogonal persistence, a pre-populated strongly typed stable store, higher-order procedures, parametric polymorphism, abstract (existential) data types, collections of name-value bindings, graphical data types, concurrent execution, two infinite union types for partial specification, and support for reflective programming. Notable additions over PS-algol include the following:

- the infinite union type *any*, which facilitates partial and incremental specification of the structure of the data
- the infinite union type *environment*, which, in addition to the above, provides dynamically extensible collections of name/L-value bindings—and thereby the dynamic construction of independent name spaces over common data
- parametric polymorphism in a style similar to that later popularised by Java generics, but with computation over truly persistent polymorphic values
- existentially quantified abstract data types for data abstraction
- a programming environment, including graphical windowing library, object browser, program editor and compiler, implemented entirely as persistent objects within the store
- support for hyper-code, in which program source code may contain embedded direct references to extant objects
- support for structural reflection, where a running program may generate new program fragments and integrate these into its own execution

The integrated persistent environment of Napier88 that supported higher-order procedures yielded a new programming paradigm, which is only possible by this means, whereby source programs could include direct links to values that already exist in the persistent environment. The programming technique was termed *hyper-programming* and the underlying representation *hyper-code*.

Hyper-code [31] is a representation of an executing system modelled as an active graph linking source code, existing values and meta-data. It unifies the concepts of source code, executable code and data, by providing a single representation (as a combination of text and hyperlinks) of software throughout its lifecycle. Sharing is represented by multiple links to the same value. Hyper-code also allows state and shared data, and thereby closure, to be preserved during evolution.

The combination of structural reflection, the ability of a program to generate new program fragments and to integrate these into its own execution, and hyper-code provides the basis for type-safe evolution. Within the persistent environment, generator programs may stop part of an executing system (while the rest of the system continues to execute), inspect its state by introspection, change the part as necessary by programming or editing the hyper-code representation, recompiling the new fragment and rebinding it into the executing system.

Unsurprisingly, given their heritage, both PS-algol and Napier88 support all three of the principles of orthogonal persistence.

## 3.4 Arjuna

The focus of the Arjuna system [32,33] is to support fault-tolerant distributed applications, based upon persistent objects supporting nested atomic actions. Atomic actions control sequences of local and remote operations against abstract datatypes implemented using C++ classes. The file system is used for long-term storage of objects. To support recoverability, a snapshot of object state is taken before an object is modified for the first time within the scope of an atomic action. This mechanism is also used to support persistence, with the new state of an object being used to replace its old state at commit time. A *state manager* provides operations to save and restore the state of object instances.

Since all persistent classes must extend the base class *StateManager*, which provides the mechanisms for persistence and atomic actions, Arjuna does not adhere to the principle of datatype orthogonality. It does not meet the requirements of persistence independence, since the programmer must implement *save_state* and *restore_state* operations for all persistent classes. Finally, for the same reason, it does not support persistence identification by reachability.

## 3.5 Persistent Java

Several orthogonally persistent versions of Java have been implemented. In PJama [34] the programmer uses an API to associate objects with strings in a persistent map in order to make them persistent. All objects transitively reachable from the map are automatically made persistent. The language syntax itself is unchanged; typically persistence can be introduced to a previously existing application with the addition of a relatively small amount of code making API calls. The compiler and standard libraries are also unchanged. The virtual machine is modified, to move objects to and from a proprietary object store automatically as required. A version of hyper-code has been prototyped using PJama [35].

The emphasis in ANU-OPJ [36] is on promoting inter-operability, by avoiding any modifications to the virtual machine. Instead, read and write barriers are introduced by dynamic byte-code modification. This is achieved by using a customised class loader, making the approach compatible with standard compilers and virtual machines. The programmer's view of persistence is slightly different from PJama, in that no persistence API is involved. Instead, all static fields are implicitly persistent roots. The Shore storage manager [37] provides object storage.

Persistent Java was implemented on the Grasshopper operating system [38]. Unlike the other persistent Java systems, no modifications were made to the abstract machine or to the bytecode generated for a particular application. Instead, orthogonal persistence was achieved by instantiating the entire Java machine within a persistent address space. In this system, like the later ANU-OPJ system, static fields were implicitly roots of persistence.

The three persistent Java systems adhered to the three principles of orthogonal persistence to varying degrees. PJama followed the PS-algol persistence model but could not make some types persistent due to restrictions in the abstract machine. Similarly, ANU-OPJ could not uniformly perform byte code transformation on some system classes. The Grasshopper version did adhere to the three principles, by virtue of making the entire environment persistent.

## 3.6 OODBs

Object-oriented database systems emerged in the mid 1980s and married persistence to object-oriented languages [39]. In the early systems, the language used tended to be an extension of C++. The Exodus System with its E programming language typified this approach [40].

The Object-Oriented Database Manifesto [41], published in 1989, set out to lay down the ground rules of what was (and what was not) an object-oriented database. It defined a number of mandatory, optional and open issues in OODB design. Space prohibits a full exposition of all the mandatory features (identity, encapsulation, computational completeness, types or classes, class hierarchies, complex objects, overriding, overloading and late binding, extensibility, persistence, secondary storage management, concurrency, recovery and ad-hoc querying); we will therefore comment on what we consider to be the most important here.

The first of these, identity, is perhaps the biggest differentiating feature between an OODB and a relational DB. Relational systems impose identity via primary keys stored as attributes, whereas objects have unique identities formed when they are created and remaining throughout their lifetimes irrespective of their states.

The issue of encapsulation is another feature that distinguishes the relational from the OO world. In a relational system the universe of discourse is made up of relations containing flat tuples, which may be queried using a relational language. By contrast, in an OO system an object has an interface, some state and a procedural component, which implements the interface and may perform operations on the state.

A last issue with OODB systems is whether code should be stored in the database; this issue seems to divide the OODB community. Many feel that putting code in the database has a detrimental effect on performance; the reasons for this are unclear. If code is not stored in the database, well-known semantic anomalies can arise. Richardson [42] describes how a program can populate a database with objects of some type *T*. Another program can insert into this data-structure an object of type *T'*, a subtype of *T*. If the original program then accesses the new object and calls methods that have been over-ridden in *T'*, it should of course use the code of the subtype when operations are performed (late binding). However, the code for *T'* may not be in the static environment (in the file system) of the original program. Indeed, the code may not even exist on the machine on which the program is written. In this case, when the original program invokes an operation on the new object a dynamic failure will result. There are essentially two solutions to this problem: relying on being able to load code from the file system—which is manifestly unsafe—or placing code in the database.

The provision of declarative querying was the primary difference between persistent languages and OODB systems; the latter generally provided querying whilst the former did not. Whilst pointer chasing can be more efficient than some operations, notably outer joins, in database systems, the inability to perform declarative queries over non-resident data is often cited as the primary reason for the lack of uptake of OODB and persistent systems. The relatively recent ability to tightly integrate query languages over objects with a host object-oriented language [3,43] has addressed much of this criticism.

Another perceived issue with OODB systems is the degree of coupling exhibited. Data in relational systems is loosely coupled; tuples are associated solely via primary and foreign key values. This permits database schemata to be refactored by database administrators independently of the code base. In an object-relational system there is also loose coupling between the code and the data. The object-relational mappings are partial; they specify a degree of compliance required of the database by the code. Thus database schema changes may not affect the code in any way. By contrast, this is not true in OODB systems, which are highly coupled in two respects: the referential integrity of pointers and type constraints specified in the programming language. Since OODB systems typically rely on being able to follow the transitive closure of objects, changes to the code and the database must be made in a consistent manner.

Most OODB systems are strongly typed and consequently the types of referends and referees must be type compliant; resulting in the schema and the code being highly coupled. A last problem perceived with OODB systems is that it is often difficult to determine the extent of pointers in the system due to lack of sufficient encapsulation. Consequently changes to the schema could affect code in arbitrary locations. However, this problem also applies to relational systems in which there is a mismatch in the integrity constraints provided by the database and those expected of the programs that compute over it. Furthermore, in a pure object-oriented system the integrity of the data may be enforced by encapsulation, which is not true in relational systems. Clearly modern software engineering tools could be brought to bear on these problems.

## 3.7 db4o

*db4o* [44] is a modern OODB system which may be used with both .NET and Java, via the provision of separate libraries for the two languages. *db4o* requires no mappings between transient and persistent data to be described by the programmer. Thus the objects stored in the database are real POJOs with no extra interfaces, extended classes or annotations. The *db4o* model is reminiscent of PS-algol. To access the database the programmer writes code such as that shown below.

```
ObjectContainer db=Db4o.openFile(Util.DB4OFILENAME);
try {
   Person al = new Person("al", 49);
   db.set(al);
}
finally {
   db.close();
}
```

The root of the database is a collection (an *ObjectSet*) of objects. It is possible to access such a persistent collection using *query by example* (QBE), by performing a *get* operation with either a prototypical object or an instance of class *Class* as a parameter. In addition, *db4o* supports both native queries and Simple Object Database Access (SODA). Native queries are constructed using predicates in C# or Java whereas SODA queries are relatively low level, using strings to select fields from objects. Once a root object has been accessed its closure may be traversed using traditional pointer following operations.

However, by default *db4o* does not load entire closures from persistent storage. *db4o* introduces a concept known as *activation depth*, which determines how much of an object closure is loaded when a parent object is loaded. By default, only the first five levels of objects are loaded from the database. It also includes mechanisms to control activation based on class, via global settings and transparently. Additionally, objects referenced from a loaded object can be loaded by explicitly activating objects as they are loaded.

To update objects stored in the database the programmer has to retrieve an object and call *set* with a top-level object as a parameter (as in the above example). However, like object loading, the entire closure of the object is not written to persistent storage on commit. Instead, the amount of closure written to storage is controlled by a concept known as *update depth* (the default is 1). Like activation depth it is possible to control update depth in a variety of ways. These design decisions have clearly been made for a mixture of implementation and efficiency reasons.

Whenever a container is opened, *db4o* implicitly starts a new transaction and an explicit commit occurs before the container is closed. A *rollback* operation permits transactions to abort. However, this operation is the root of a semantic anomaly. Loaded instances of database objects may be still be accessible yet out-of-sync with the store. To address this problem *db4o* provides a *refresh* operation, which may be applied to objects. It is unclear how the programmer is supposed to know which objects require refreshing; again this deviates from the principles of orthogonal persistence.

The *db4o* system adheres to the principle of data independence. No mappings or annotations are required to indicate which types may be made persistent. Similarly, code may manipulate data independent of its longevity. The concepts of update and activation depth do impact this principle since, for example, a method to determine the length of a list might get the wrong answer if activation depth was not used correctly. This is seen as desirable by the developers who state that "*db4o provides a mechanism to give the client fine-grained control over how much he wants to pull out of the database when asking for an object*" [45]. This property seems not to preserve identity. The principle of data type orthogonality is adhered to, since any user-defined data object can be made persistent without any additional code, annotations or XML specifications.

## 3.8 Java Data Objects

Java Data Objects was released in 2002 [46], providing a storage interface for Java objects without the necessity to interact with data access languages such as SQL. Using JDO, Java objects may be stored in a relational database, an object database, XML file, or any other technology using the same interface. Since it enables Java programmers to transparently access underlying data storage without using database-specific code, it moves considerably towards the goals of persistent systems. An example of the use of JDO is shown below. Although not shown in this example, the entire transitive closure of objects is stored in the database on commit.

```
PersistenceManagerFactory pmf = JDOHelper.
   getPersistenceManagerFactory(..);
PersistenceManager pm = pmf.getPersistenceManager();
Person p = new Person("Bob Smith", 49 );
Transaction tx;
try {
  tx = pm.currentTransaction();
  tx.begin();
  pm.makePersistent(p);
  tx.commit();
} catch (Exception e) { … }
```

Although this looks very much like the PS-algol examples, much additional specification is required when using JDO. The relationship between the Java objects and persistent data is specified using an XML metadata file. A simple example is shown below, specifying the persistent class *com.xyz.Person*. Field modifiers may specify a number of attributes, including which fields are primary keys, whether fields are persistent or transient, how fields are to be loaded, and how null values should be handled.

```
<?xml version="1.0" encoding="UTF-8"?>
<!DOCTYPE jdo SYSTEM "jdo.dtd">
<jdo>
   <package name="com.xyz">
      <class name="Person">
         <field name="firstname"
         persistence-modifier="persistent"/>
         …
      </class>
   </package>
</jdo>
```

The query language provided by JDO, JDO Query Language (JDOQL), abstracts over the underlying storage technology. A query interface selects objects from the database irrespective of whether the underlying storage is based on objects or relations. Queries are passed to the persistence manager and operate on either class extents or explicit collections. Filtering is provided by providing Boolean expressions which are applied to instances.

```
Query query = pm.newQuery(Person.class, people,
    "name == \"Malcolm Atkinson\"");
Collection result = (Collection) query.execute();
Iterator iter = result.iterator();
while ( iter.hasNext() ) {
    Person p = (Person) iter.next();
    ....
}
```

JDO succeeds in abstracting over particular underlying storage technologies. However, in some cases, notably relational databases, the mapping between language objects and storage level objects must be described. When an object-relational mapping is used with JDO, the O-R mappings are described in ORM mapping files.

Persistence of data is independent of the programs manipulating it, provided that appropriate persistence mappings have been described. The principle of data type orthogonality is violated, since only those objects that have a persistent mapping can be made persistent. Furthermore, system classes and some collection classes may not be made persistent.

## 3.9 Java Persistence API

The Java Persistence API [47] is intended to operate inside or outside a J2EE container, creating a persistence model for (plain old) Java objects. It eliminates much of the complexity required by JDO. For example, the XML mapping tables are no longer required, and the objects that can be made persistent are ordinary Java objects rather than having to implement specified interfaces. In contrast to JDO, which is agnostic to storage technology, the Java Persistence API is explicitly for use in an object-relational context.

```
@Entity
public class Person
    public Person() {}
    @column( name="name" )
    public String getName() {}
    @column( name="age" )
    public int getAge() {}
}
```

The *@Enitity* annotation can be decorated with parameters specifying the name of the table from which data is drawn; by default this is the name of the class. Similarly, the column name may be specified using the *@column* annotation and identity attribute using *@id*. This is clearly not a POJO system, despite being often described as one, since it requires annotations to be made in the Java classes describing the object-relational mappings. Object-relational mappings can be arbitrarily complex, and it is possible to specify that data be drawn from multiple tables using join-based queries.

Queries are defined using (an extension to) Enterprise JavaBeans query language (EJB QL) rather than SQL. The difference is subtle but important: rather than querying over tables in the database, queries are performed on the beans and the relationships between them. These relationships are specified using the attributes embedded within the Java objects.

Using the Java Persistence API the persistence of data is independent of the programs that manipulate data. Additionally, the programmer does not have any explicit control over the movement of data between the store and main memory, thus adhering to the principle of persistence independence. The principle of data type orthogonality is only partially adhered to, since only instances of classes that are decorated with an *@Entity* annotation may be stored in the persistent store. This explicitly precludes most system classes from being persistently stored. The principle of persistence identification is largely adhered, to since the mechanism for identifying persistent objects is not related to the type system.

Despite not being fully compliant with the principles of orthogonal persistence, an application programmer can program against persistent data without the knowledge that the data is persistent. This is very much in the spirit of the aims of orthogonal persistence.

## 3.10 LINQ

Microsoft, recognizing the problems of embedding queries into programs as strings, has created Language-Integrated Query (LINQ) [3]. Unlike the Java systems described previously, the approach taken by LINQ is to add general-purpose query facilities that may be applied to all information sources. Thus being able to query over relational data is merely a special case of querying. For example, using LINQ it is possible to write a C# program to query over a collection of persons as follows:

```
static void doquery( Person[] people ) {
    IEnumerable<Address> result = from p in people
                                  where p.age == 49
                                  select p.address;
    foreach (Address item in result)
      Console.WriteLine(item.getTown());
}
```

The query selects all the people from the array whose age is 49 and forms an enumeration containing their addresses. Note that the query is integrated with the programming language, making it amenable to static type checking, optimization and—perhaps more importantly—design tools such as refactoring tools.

Relational data stored in a database can also be manipulated using a Visual Studio component called *LINQ to SQL*, which transparently translates LINQ queries into SQL for execution by the database engine. The results are returned in the language level objects defined in the user program. LINQ tracks the relationships between the language objects and the database transparently.

Like the Java object-relational mapping solutions, objects may be labelled with annotations to identify how properties correspond to database columns. Tool support is provided to assist in the translation between extant databases and language level object definitions.

## 4 Taking Stock

A selection of approaches to programming with persistent data have been outlined. They differ in a number of key attributes, including:

- data-centric or program-centric
- degree of adherence to the principles of orthogonal persistence
- degree of impedance mismatch
- storage technology employed
- whether object identity is automatically preserved
- whether code is stored with data
- support for declarative queries over non-resident data
- support for transactions

Space precludes a full analysis of the various approaches with respect to all of these aspects, but we suggest that the most fundamental is the overall system philosophy. In a data-centric approach it is assumed that pre-existing persistent data is a given, and the issue is how to program over that data. In a program-centric approach, code comes first, and the issue is to provide persistence of program data between executions.

In a data-centric approach the existing data is likely to be large and long-lived, and openness of the data—avoiding lock-in to proprietary technology—is likely to be important. Relational databases have overwhelming advantages in this sector: mature technology resulting from long-term investment in scalability and optimization; widely available expertise; and standard interfaces promoting inter-operability. Approaches in this category include low-level database APIs such as JDBC, and the various object-relational mapping technologies. The constraints imposed by the requirement to inter-operate with existing data—and to cope with changes to both data and meta-data made via other routes to the data—mean that none of these approaches achieve data type orthogonality, and that all involve a significant impedance mismatch. The ORM systems require the programmer to understand and specify the mapping between multiple representations, while low-level APIs also require conversion code to be written.

Designers of program-centric persistence technologies are less constrained in their choice of storage format since they may legitimately assume that the persistent data will be solely accessed via the language infrastructure. The systems that adhere to the principles of orthogonal persistence have all used proprietary closed storage formats. There is no obvious technical reason why this is a necessary choice, although it may well maximise scope for achieving good performance. This may have been one factor behind the lack of commercial adoption of the various successful research prototypes. To invest in significant use of any closed storage system requires a very high level of trust in the long-term viability of the technology and the processes that support it. Other obvious limiting factors are the relatively limited scalability of those systems in terms of size and query performance, inevitable given the resources available.

Object-relational systems have been highly successful, now dominating the field in large applications. It is clear, however, that significant impedance mismatch problems remain. Although the modern programmer is less likely to have to program the transfer of objects to and from long-term storage, they must still deal with a bewildering level of complexity in specifying mappings between objects and relations. The recent emergence of conceptually simpler approaches such as *db4o* is a sign that significant demand remains for the benefits pursued in the original investigations of orthogonal persistence.

It is perhaps also worth reflecting on the current usefulness of the principles of orthogonal persistence, a quarter century after they were first proposed. The principle of persistence independence suggests that data manipulation should be coded in exactly the same way for transient and persistent data, and that the programmer should not have to control data movement between transient and persistent storage. So long as the language is sufficiently rich that all desired data manipulation can be expressed conveniently, there seems no obvious argument against this principle. Of course, adherence to it incurs some implementation effort, hence not all approaches do so.

The principle of data type orthogonality suggests that all objects should be permitted the full range of persistence. Again, as a desirable feature this seems uncontroversial. Again, it raises significant implementation difficulties, leading to few systems achieving full adherence. Even those that claim full orthogonality have tended to have difficulty with objects that depend on external state, such as file descriptors, GUI elements, network channels etc.

The principle of persistence identification has had a more chequered history. The wording of its definition earlier is taken from [21]. In the earlier [1], however, which first proposed principles of orthogonal persistence, the principle is listed but not named. In hindsight, it now seems unclear what, precisely, is mandated by this principle that is not already covered by the principle of persistence independence. This appears to have been recognised in more recent discussion, in which it has been replaced by the more concrete principles of transitive persistence [34] and persistence by reachability [48]. We may perhaps conclude that a more useful general principle might be that it should be possible to identify persistent objects in a convenient way. If doing so via the type system is forbidden by the principle of data type orthogonality, and identifying each object individually is ruled out as too arduous, then persistence by reachability is the only obvious solution.

## 5 Future Directions

Orthogonally persistent systems will not replace object-relational systems in the foreseeable future. We may, however, speculate on niche areas in which the principles of orthogonal persistence might be usefully carried forward. One possibility is the development of a program-centric approach in which fully orthogonal persistence is implemented using a relational database as the storage engine. This would address the 'closed data format' criticism potentially levelled at previous implementations, since read-only access to the data could be permitted at the relational level.

Another potential avenue for development is to target emerging application styles such as cloud applications. The development of such applications could be significantly simplified by a system supporting programming over resilient distributed objects in a transparent manner, abstracting over replication and physical location in the same way that orthogonal persistence abstracts over storage hierarchy [49].

Another avenue for investigation is how the unique features of orthogonally persistent systems may be exploited to improve current software development technology [18]. For example, the integration of first-class code and data within a persistent store that enforces referential integrity makes the hyper-code paradigm possible. This could be extended with more sophisticated support for application system evolution, analogous to refactoring tools provided by modern IDEs [50]. Hyper-code allows source code to be reliably associated with all code objects. Thus, whereas refactoring tools currently operate separately on a code base or on a database, refactoring within a persistent environment could be applied uniformly to data and the code that operates on it. Evolutionary code could reflect over all of the data bound into the code-base being evolved, as well as the structure of the code-base itself. Arbitrary evolution (or refactoring) of a running application could be performed with complete confidence that all code and data affected by a change could be located and evolved in turn consistently. This would be possible even for data that in conventional systems would be encapsulated within closures and thus inaccessible to evolution code.

## 6 Conclusions

Orthogonal persistence was proposed to address the impedance mismatch problem. This problem has been with us for 20-30 years and refuses to go away. It has recently been described as the *Vietnam of Computer Science* [51]. Far from being resolved, the impedance mismatch is perhaps getting worse. We now have impedance mismatch across the multiple subsystems concerned with data replication, cache-coherency and distribution. In many of today's enterprise systems the programmer must, by necessity, not only manage mappings from the language to the database but also from the language to the Memcached [52] or DBCache [53] layers, and from those layers to the database. Thus, when we consider the impedance mismatch problem in our systems it is important to recognise that the object-relational mapping is not the only mapping that must be considered. Even if non-relational storage is used, for example Amazon S3 [54], mapping between layers is required. The essential issues are who creates the mappings and how efficiently they can be maintained.

In [19] Maier stated that one of the major problems of OO systems was the lack of integration between bulk operations and the programming language. In this domain good progress has been made in the last few years. LINQ makes great strides in providing a single (sub-) language that operates over objects regardless of their longevity.

The solutions to providing persistence in programming systems have been many, and the road has been long and winding. However, there has been a clear trend towards the ideals of orthogonal persistence. The state of the art has finally moved away from strings containing embedded queries with explicit coercions to values in the programming language space.

In the 1980s orthogonal persistence focussed on the differences between long- and short-term storage. As described above, this is just one of many mappings that an application builder needs to be concerned with; there are many subsystems that require mappings to be maintained, including caching, networks, virtualized hosts, distributed storage, and replication. Furthermore, we are moving towards a world in which applications are self-organising and autonomic. Such autonomic systems are likely to be concerned with data clustering, machine utilisation and the ability to distribute computation and storage. Lastly the scale of application systems is likely to vary enormously from small persistent applications on devices such as iPhones through to extremely large ones to address the scientific challenges of tomorrow. In such a world it seems unlikely that the intellectual burden of managing a plethora of complex mappings can be left in the human domain.

## 7 Acknowledgements

Our experience in the design and implementation of persistence systems has benefitted from interaction with so many people that it would be invidious to mention a recently remembered subset. Malcolm Atkinson deserves mention as the inventor of the persistence concept and we would like to thank the community that populated the Persistent Object Systems (POS) and Database Programming Language (DBPL) Workshops where much of this work was reported and digested.